\documentclass[12pt]{iopart}
\usepackage[dvips]{graphicx}
\begin{document}
\title{Magnetopumping Current in Graphene Corbino Pump}
\author{Babak Abdollahipour\footnote{Emails: b-abdollahi@tabrizu.ac.ir}}
\address{Faculty of Physics, University of Tabriz, Tabriz 51666-16471, Iran}
\author{Elham Moomivand}
\address{Department of Physics, Faculty of Science, University of Zanjan, Zanjan 45371-38791, Iran}
\begin{abstract}
We study conductance and adiabatic pumped charge and spin currents in a graphene quantum pump with Corbino
geometry in the presence of an applied perpendicular magnetic field. The pump is driven by the periodic and
out of phase modulations of the magnetic field and an electrostatic potential applied to the ring area of the
pump. We show that the Zeeman splitting, despite of its smallness, can suppress the conductance oscillations
at the zero doping and in a threshold value for the flux piercing the ring area which depends on the inner
lead radius and thus on the flux penetrating in it. Moreover, it generates a considerable spin conductance at
infinitesimal nonzero doping and at the magnetic flux, that charge conductance starts to suppress. We find
that the pumped charge and spin currents increase by the magnetic field with small oscillations until they
start to suppress due to the effect of the nonzero doping and the Zeeman splitting. In graphene Corbino pumps
with small inner leads the Zeeman splitting shows its effect in a large value of the magnetic field and thus
we can get a considerable pumped charge and spin currents at the enough small magnetic fields.
\end{abstract}
\maketitle

%%%%%%%%%%%%%%%%%%%%%%%%%%%%%%%%%%%%%%%%%%%%%%%%%%%%%%%%%%%%%%%%%%%%%
\section{Introduction}
%%%%%%%%%%%%%%%%%%%%%%%%%%%%%%%%%%%%%%%%%%%%%%%%%%%%%%%%%%%%%%%%%%%%%
Since the advent of graphene it has been the subject of intense theoretical and experimental research mainly
due to its very peculiar electronic structure. Low energy excitations in graphene are well described by the
Dirac-like equation in contrast to normal semiconductors with quadratic dispersion law. Charge carriers in
graphene being massless possesses the unique feature of chirality which leads to several exotic transport
behaviors such as the minimum of the conductivity\cite{Novoselov05}, the Klein tunnelling\cite{Katsnelson06},
half-integer quantum Hall effect\cite{Novoselov06} and many other interesting effects\cite{Neto09}. Moreover,
owing to a very long spin relaxation time, which is expected due to a very weak intrinsic spin-orbit
interaction, and the extrinsic spin-orbit coupling appearing in the presence of a transverse electric field,
graphene is also attractive for applications in spintronics\cite{Konschuh10}. Many theoretical and
experimental works have been devoted to study injection and manipulation of the spin in graphene\cite{Han14}.

A distinct feature of a wide and short strip of undoped graphene is that it exhibits transport properties
which are indistinguishable from those of a classical diffusive systems\cite{Tworzydlo06}. It has been shown
that in this regime, the so called pseudo-diffusive regime, conductivity and the higher current cumulants are
not affected by the applied magnetic field\cite{Ostrovsky06,Louis07,Prada07}. These effects are well explained
by the evanescent modes which are present at the Dirac point and transport entirely occurs via them with a
transmission that is equal to the diffusive transport theory result\cite{Akhmerov07}. At sufficiently strong
magnetic fields an exponentially small chemical potential is enough to enter a field-suppressed transport
regime. However, at resonance with the Landau levels the pseudo-diffusive behavior is recovered for all
current cumulants\cite{Prada07}. In a collection of quantum billiards of undoped graphene, including Corbino
disk, the pseudo-diffusive behavior has been identified by Rycerz {\it et al.}\cite{Rycerz09}. Moreover, they
showed that by shrinking at last one of the billiard openings a crossover from the pseudo-diffusive to the
quantum-tunneling regime is observed. Rycerz has investigated the effect of the magnetic filed on the
conductance, the Fano factor and the higher current cumulants of the Corbino disk in graphene. He showed that
conductance of the graphene Corbino disk is a periodic function of the magnetic flux piercing the disk area at
zero doping\cite{Rycerz10}. Also, this effect is followed by the shot noise and higher current
cumulants\cite{Rut15}. Moreover, it has been found the periodic magnetoconductance oscillations decay rapidly
with increasing field at finite voltage in contrast to the higher cumulants which show periodic oscillations
for arbitrary high fields. The periodic oscillations of the conductance as a function of the magnetic flux
piercing the disk has been found to persist in the bilayer grahpene\cite{Rut14}.

Most of researches about the electronic transport in graphene has been devoted to stationary problems.
Non-stationary problems may result in emergence of variety of new effects. Quantum pumping is one of these
interesting phenomenon, initially due to D. J. Thouless\cite{Thouless83} and then developed by several
successive studies\cite{Brouwer98,Zhou99,Moskalets02}, in which a periodic modulation of parameters of a
nanoscale system produces a finite dc current through it even in the absence of an external bias. It has been
suggested to exploit and probe the properties of graphene both by adiabatic and non-adiabatic quantum
pumping\cite{Zhu09,Prada09,Zhu11,Abdollahipour14,Grichuk10,Wang10,Tiwari10,Zhang11,Zhang12,Zhang12JPCM,Palop11,SanJose11,Wu12}.
In the adiabatic regime, when the pumping cycle is very slower than the carrier passage from the pump region,
for generating a pumped current at last two system parameters should vary periodically and out of
phase\cite{Brouwer98}. Studies on graphene have given rise to new insights into quantum pumping in graphene.
Pumped current in graphene shows an increment as compared with semiconductor materials, which is attributed to
the unusual electronic spectrum of graphene and in particular, the important role of the Klein tunneling and
evanescent modes at the Dirac point\cite{Prada09,Zhu11,Abdollahipour14}. In several works, the quantum pump
effect has been proposed for potential use in graphene-based
spintronics\cite{Wang10,Tiwari10,Zhang11,Zhang12,Zhang12JPCM} and
vallytronics\cite{Wang14,Wang14-2,Mohammadkhani15} as an effective tool for generating and controlling spin
and valley currents. Graphene pumps therefore offer a number of potential advantages which may prove useful
for practical applications. Envisaged applications for graphene pumps outside the valleytronics and
spintronics include quantum metrology, single photon generation via electron-hole recombination in
electrostatically doped bilayer graphene reservoirs\cite{Mueller09}, and for read-out of spin-based graphene
qubits in quantum information processing\cite{Trauzettel07}. Attempts have been made to experimentally
investigate adiabatic quantum pumping by using electrostatically defined quantum dots in GaAs-based
heterostructures\cite{Switkes99}. In such a system, pumping is induced by oscillating voltages applied to the
gate electrodes defining the dot. In these experiments, the resultant pumped current are affected by stray
bias voltages originating from parasitic coupling of the ac signals applied to the
gates\cite{Brouwer01,DiCarlo03}. Recently quantum pumping in graphene has been realized in the graphene double
quantum dot pump driving by two oscillating gate potentials which control the total number of electrons on
each quantum dot\cite{Connolly13}.

In the studies on magnetoconductance in graphene the spin degree of freedom is usually ignored. Despite the
smallness of the Zeeman splitting, it has been shown that it can show remarkable effects on the electronic
transport in graphene such as the appearance of the chiral spin edge states\cite{Abanin06}, giant spin-Hall
effect in the absence of the spin-orbit interaction\cite{Abanin11} and generating a large spin pumped
current\cite{Tiwari10}. Motivated by the recent researches about the quantum transport in graphene Corbino
disk, in this paper we first analyze effect of the zeeman splitting on the magnetoconductance, totally
disregarded in the previous studies, and we also study the quantum pumping in graphene Corbino pump driven by
periodic and out of phase modulation of an electrostatic potential and a magnetic field both applied to disk
area of the system. We find that the Zeeman splitting, despite its smallness, suppresses the periodic
oscillations of the magnetoconductance at the Dirac point. Moreover, it produces a considerable spin
conductance about the flux threshold value, where the magnetoconductance starts to suppress, in low magnetic
fields and at the resonance with the Landaue levels in the high magnetic fields. At the Dirac point the pumped
charge and spin currents increase as a function of the magnetic flux with small periodic oscillations until
they suppress by the effect of the Zeeman splitting. For nonzero doping the effects of the doping and the
zeeman splitting compte with each other in suppressing the pumped current. Growing nature of the pumped
currents by the magnetic flux provides the possibility of the producing a large pumped currents at low
magnetic fields and it is very striking that it dose not suffer the stray potentials problem. Moreover, we
have a considerable charge and spin pumped currents at the resonances mediated by the Landaue levels which
provides an extra control on the charge and spin current via changing the doping level by the potential
applied to the gates.

The outline of the paper is as follow. In section \ref{model} we introduce our model and the procedure for
calculating of the magnetoconductance and magnetopumping. Section \ref{results} is devoted to explaining our
main results and discussions concerning them. Finally, we give conclusion of our work in section
\ref{conclusion}.
%
%%%%%%%%%%%%%%%%%%%%%%%%%%%%%%%%%%%%%%%%%%%%%%%%%%%%%%%%%%%%%%%%%%%%%
\section{Model and basic equations}\label{model}
%%%%%%%%%%%%%%%%%%%%%%%%%%%%%%%%%%%%%%%%%%%%%%%%%%%%%%%%%%%%%%%%%%%%%
%
%%%%%%%%%%%%%%%%%%%%%%%%%%%%%%%%%%%%%%%%%%%%%%%%%%%%%%%%%%%%%%%%%%%%
%%%%%%%%%%%%%%%%%%%%%      Fig 1     %%%%%%%%%%%%%%%%%%%%%%%%%%%%%%%
\begin{figure}[t]
\centerline{\includegraphics[width=7cm]{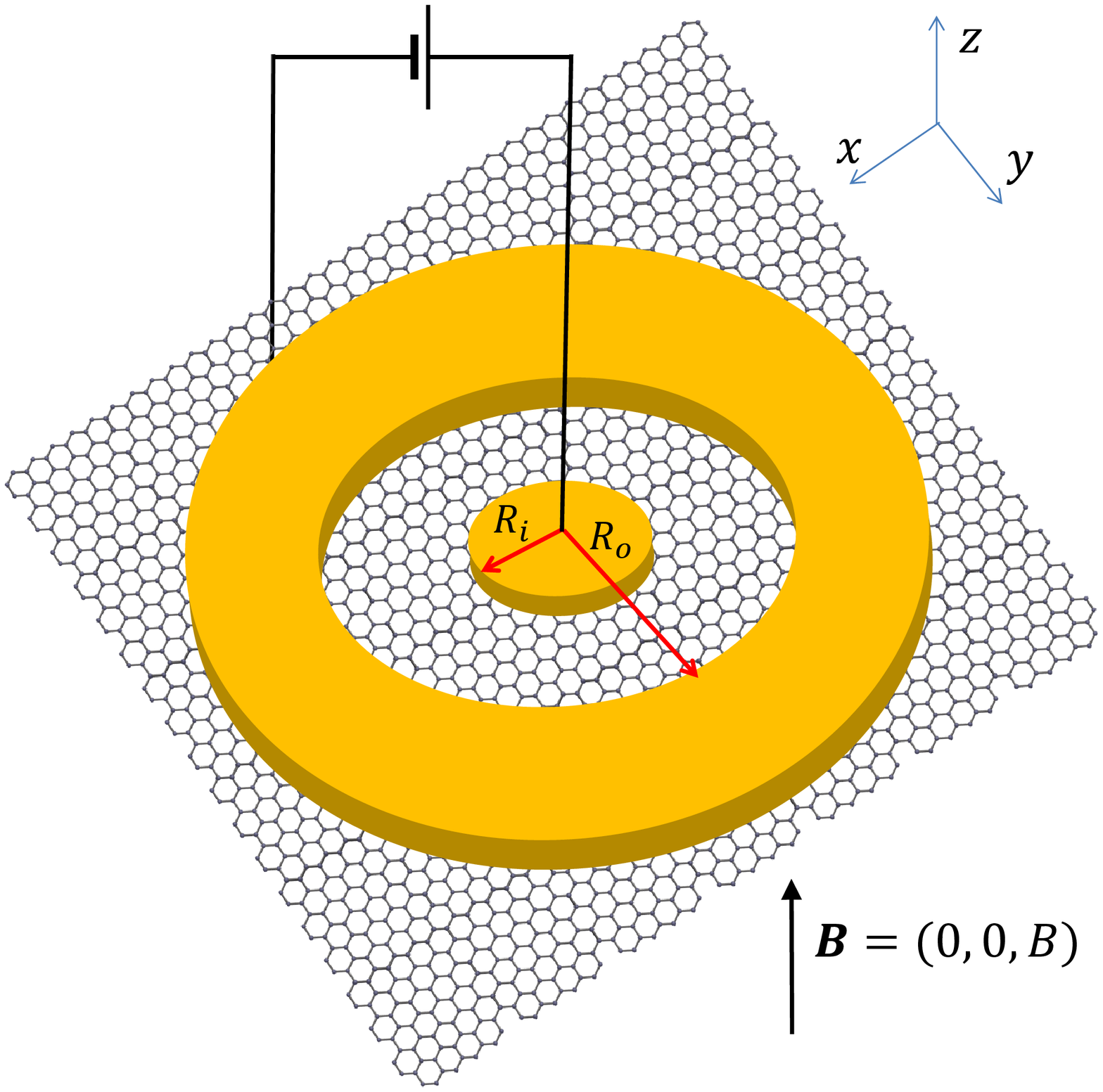}} \caption{(Color online) Graphene Corbino quantum pump is
composed of a disk shaped area with the inner radius $R_{i}$ and outer radius $R_{o}$ connected to the
metallic leads. Pumping current is generated by the periodic and out of phase variations of an electrostatic
potential $U$ exerted via a back gate and a perpendicular magnetic field $B$ applied to the pump.}
\label{system}
\end{figure}
%%%%%%%%%%%%%%%%%%%%%%%%%%%%%%%%%%%%%%%%%%%%%%%%%%%%%%%%%%%%%%%%%%%%
%

We consider a graphene based quantum pump with Corbino disk geometry in which a disk shaped graphene is
surrounded from both exterior and interior sides with circular metallic leads. It has been schematically
depicted in Fig.\ref{system}. Advantage of the Corbino geometry in comparison to the planar structures is in
the absence of the edge boundary. The metallic leads are modeled by the heavily doped graphene regions.
Pumping current through the system is generated by the periodic and out of phase variations of an
electrostatic potential $U$ exerted via a back gate and an external magnetic field $B$ applied perpendicular
to the graphene sheet. The periodic variation of the pumping parameters is give by
\begin{equation}\nonumber
B(t)=B_{0}+\delta B \cos(\omega t)\ ,\\
U(t)=U_{0}+\delta U\cos(\omega t +\phi)\ ,
\end{equation}
where $\delta B$ and $\delta U$ are amplitudes of variations about the equilibrium state values $B_{0}$ and
$U_{0}$, respectively. In the above equation $\phi$ is the phase difference between variations of two
parameters and play the important role in current pumping. In the absence of the processes which mixes two
spin states we can treat each spin state separately. In the bilinear response regime, where $\delta B\ll
B_{0}$ and $\delta U\ll U_{0}$, the adiabatic pumped current for electrons with spin $\gamma(=+,-)$ is given
by the Brouwer's formula which is based on the scattering approach \cite{Brouwer98}:
\begin{equation}\label{Brouwer-formula}
I^{\gamma}_{p}=I_{0}\sum_{\alpha \in L,R} Im\left(\frac{\partial S^{\gamma}_{L,\alpha}} {\partial U}
\frac{\partial (S^{\gamma}_{L,\alpha})^\ast}{\partial B}\right)\ ,
\end{equation}
where $I_{0}=\frac{\omega}{2\pi}e\delta U\delta B\sin(\phi)$ and $S^{\gamma}_{L,\alpha}$ is an element of the
scattering matrix for spin $\gamma$ electron. In the above equation summation goes over all of the modes in
the left ($L$) and right ($R$) leads and the symbol "$Im$" denotes the imaginary part. Pumped charge and spin
currents are given by $I^{c}_p=I^{+}_{p}+I^{-}_{p}$ and $I^s_p=I^{+}_{p}-I^{-}_{p}$, respectively. The
adiabatic limit for pumping current is satisfied when oscillation period of the driving parameters be much
longer than the dwell time of the carriers in the system, namely $\tau_{dw}\ll 2 \pi/\omega$.

We use scattering approach to calculate the scattering matrix of the pump. The low energy excitations for
$\gamma$ spin carriers in graphene in the presence of an applied magnetic filed are described by
two-dimensional Dirac equation
\begin{equation}
H_{\gamma}=v_{F}(\mathbf{p}+e\mathbf{A})\cdot\mbox{\boldmath $\sigma$} +\gamma E_z + U(r)\ ,
\label{Hamiltonian}
\end{equation}
where $v_{F}$ is Fermi velocity, $\mathbf{p}=(p_{x},p_{y})$ is the momentum operator and $\mbox{\boldmath
$\sigma$}=(\sigma_{x},\sigma_{y})$ is the vector of the Pauli matrices. We choose the symmetric gauge
$\mathbf{A}=B/2(-y,x)$, where the magnetic field $B$ is applied perpendicular to the graphene sheet. First
term of Eq. \ref{Hamiltonian} denotes effect of the magnetic field on momentum and second term is the Zeeman
splitting with $E_z=\mu_BB/2$, where $\mu_B$ is the Bohr magneton. The electrostatic potential energy is given
by $ U(r)=U\Theta(r-R_i)\Theta(-r+R_o)-U_{\infty}(-r+R_i)\Theta(r-R_o)$, where $U$ and $U_{\infty}$ are
electrostatic potentials in the disk area and the leads. We follow Ref. \cite{Rycerz10} to obtain the
eigenvectors in the presence of the magnetic field. Since the Hamiltonian given by Eq. \ref{Hamiltonian}
commutes with the total angular momentum operator $J_{z}=-i\hbar\partial_{\varphi} + \hbar\sigma_{z}/2$, thus
the energy eigenvectors must be the simultaneous eigenstates of $J_{z}$. Then, in the polar coordinates
$(r,\varphi)$ we can suppose the following form for the eigenvectors,
\begin{eqnarray}
\psi^{\gamma}_{j}(r,\varphi)=e^{i(j-1/2)\varphi} \left(
\begin{array}{cc}
\begin{array}{c}
\chi _{j\uparrow}(r)
\end{array}
\\*
\begin{array}{c}
e^{i\varphi}\chi _{j\downarrow}(r)
\end{array}
\end{array}
\right)\ ,
\end{eqnarray}
where $j$ is a half-odd integer and the pseudospin is denoted by $ s=\uparrow,\downarrow $. The radial
component $\chi_{j}(r)=[\chi_{j\uparrow}(r),\chi_{j\downarrow}(r)]^{T}$ satisfies $H^{\gamma}_j(r)\chi_j
(r)=E\chi_j(r)$, where
\begin{eqnarray}
H^{\gamma}_{j}(r)&=&-i\hbar v_{F}\sigma_{x}\partial_r+\gamma E_z+U(r)+\nonumber\\
&&\frac{\hbar v_{F} \sigma_{x}}{r} \left(\begin{array}{cc}
\begin{array}{c}
(j-1/2)+\frac{e B r^{2}}{2\hbar}
\end{array}
&
\begin{array}{c}
0
\end{array}
\\*
\begin{array}{c}
0
\end{array}
&
\begin{array}{c}
(j+1/2)+\frac{e B r^{2}}{2\hbar}
\end{array}
\end{array}
\right).
\end{eqnarray}
Now, it is possible to separately solve the scattering problem for each $j-th$ angular momentum eigenstate
incoming from the origin $ (r=0)$. This reduces the full problem to an effective one-dimensional scattering
problem for the spinor $\chi_{j}(r)$. The normal leads are modeled by the heavily-doped graphene regions by
taking the limit of $U(r)=U_{\infty}\rightarrow\mp\infty$ (the minus sign refers to the conduction band, and
the positive sign refers to the valence band). We can write the wavefunctions in the different regions as
following. In the inner lead $(r < R_{i})$, it is
\begin{equation} \chi_{j}^{(i)}=\frac{e^{ik_{\infty}r}}{\sqrt{r}}
\left(
\begin{array}{cc}
\begin{array}{c}
1
\end{array}
\\*
\begin{array}{c}
1
\end{array}
\end{array}
\right)+r^{\gamma}_{j} \frac{e^{\mp i k_{\infty }r}}{\sqrt{r}} \left(
\begin{array}{cc}
\begin{array}{c}
1
\end{array}
\\*
\begin{array}{c}
-1
\end{array}
\end{array}
\right)\ ,
\end{equation}
where $r^{\gamma}_{j}$ is reflection coefficient in the $j$th mode. In the outer lead $(r > R_o)$ the
wavefunction is given by
\begin{equation}
\chi_{j}^{(o)}=t^{\gamma}_{j}\frac{e^{\pm ik_{\infty}r}}{\sqrt{r}} \left(
\begin{array}{cc}
\begin{array}{c}
1
\end{array}
\\*
\begin{array}{c}
1
\end{array}
\end{array}
\right)\ ,
\end{equation}
where $t^{\gamma}_j$ represents transmission coefficient in the $j$th mode. In the above equations we have
introduced $k_{\infty}\equiv \mid E-U_{\infty}\mid/\hbar v_{F}\rightarrow\infty$. Finally, wavefunction in the
pumping region ($R_i<r<R_o$) subjected to a perpendicular magnetic field is given by\cite{Rycerz10}
\begin{equation}
\chi_{j}^{(d)}=A_{j} \left(
\begin{array}{cc}
\begin{array}{c}
\xi_{j\uparrow}^{(1)}
\end{array}
\\*
\begin{array}{c}
\pm i z_{j,1}\xi_{j\downarrow}^{(1)}
\end{array}
\end{array}
\right) +B_{j} \left(
\begin{array}{cc}
\begin{array}{c}
\xi_{j\uparrow}^{(2)}
\end{array}
\\*
\begin{array}{c}
\pm i z_{j,2}\xi_{j\downarrow}^{(2)}
\end{array}
\end{array}
\right)\ ,
\end{equation}
where $z_{j,1}=[2(j+s_{j})]^{-2s_{j}}$ and $z_{j,2}= 2(\frac{\beta}{(k^{\gamma})^2})^{s_{j}+\frac{1}{2}}$ with
$ s_{j}\equiv \frac{1}{2} sign(j)$, $\beta=(\frac{eB}{2\hbar})$ and $ k^{\gamma}\equiv \mid E-U-\gamma E_z\mid
/(\hbar v_{F})$. The other parameters are given by
\begin{equation}
\xi_{js}^{(\nu)}=e^{-\beta r^{2}/2}(k^{\gamma}r)^{\mid L_{s}\mid} \left\{
\begin{array}{cl}
M(\alpha_{js},\gamma_{js},\beta r^{2}) \  ,  if \quad  \nu =1,
\\
U(\alpha_{js},\gamma_{js},\beta r^{2}) \ \ , if \quad \nu =2,
\end{array}\right.
\end{equation}
where $\alpha_{js}=\frac{1}{4}[2(j+m_{s}+|j-m_{s}|+1)-\frac{(k^{\gamma})^2}{\beta}]$, and $\gamma_{js}=|j
-m_{s}|+1$ with $m_{s}=\pm\frac{1}{2}$ for the pseudospin $ s=\uparrow ,\downarrow $. The functions $M(a,b,z)
$ and $U(a,b,z)$ are the confluent hypergeometric functions\cite{Abramowitz}. By applying mode matching at the
boundaries $ \chi_{j}^{(i)}(R_{i})=\chi_{j}^{(d)}(R_{i}) $ and $\chi_{j}^{(o)}(R_{o})=\chi_{j}^{(d)}(R_{o}) $,
we can find the reflection and transmission coefficients as a function of the magnetic flux $\phi/\phi_{0}$,
where $ \phi=\pi (R_{o}^{2}-R_{i}^{2})B $ and $ \phi_{0}=2(\frac{h}{e}) Ln(\frac{R_{o}}{R_{i}})$, as
following,
\begin{eqnarray}
r^{\gamma}_{j}&=&e^{2iR_ik_{\infty}}
\\&&\nonumber
\frac{-z_{j,1}\xi^{(1)}_{j,\downarrow}(R_o)\eta^{(2)}_{j,+}(R_i)+
i\xi^{(1)}_{j,\uparrow}(R_o)\eta^{(2)}_{j,+}(R_i)+\eta^{(1)}_{j,+}(R_i)\eta^{(2)}_{j,+}(R_o)}
{z_{j,1}\xi^{(1)}_{j,\downarrow}(R_o)\eta^{(2)}_{j,-}(R_i)+i\xi^{(1)}_{j,\uparrow}(R_o)\eta^{(2)}_{j,-}(R_i)
-\eta^{(1)}_{j,-}(R_i)\eta^{(2)}_{j,-}(R_0)}\ ,
\\
t^{\gamma}_{j}&=&2i\sqrt{\frac{R_o}{R_i}}e^{i(R_i-R_o)k_{\infty}}
\\&&\nonumber
\frac{z_{j,2}\xi^{(1)}_{j,\uparrow}(R_o)
\xi^{(2)}_{j,\downarrow}(R_o)-z_{j,1}\xi^{(1)}_{j,\downarrow}(R_o)\xi^{(2)}_{j,\uparrow}(R_o)}
{z_{j,1}\xi^{(1)}_{j,\downarrow}(R_o)\eta^{(2)}_{j,-}(R_i)+i\xi^{(1)}_{j,\uparrow}(R_o)\eta^{(2)}_{j,-}(R_i)
-\eta^{(1)}_{j,-}(R_i)\eta^{(2)}_{j,-}(R_i)}\ ,
\end{eqnarray}
where
\begin{eqnarray}
\eta^{(i)}_{j,\pm}(R_i)=z_{j,i}\xi^{(i)}_{j,\downarrow}(R_i)\pm i\xi^{(i)}_{j,\uparrow}(R_i)\ ,
\\
\eta^{(i)}_{j,\pm}(R_o)=z_{j,i}\xi^{(i)}_{j,\downarrow}(R_o)\pm i\xi^{(i)}_{j,\uparrow}(R_o)\ .
\end{eqnarray}
Charge and spin conductances are defined by $G_{c}=G^{+}+G^{-}$ and $G_{s}=G^{+}-G^{-}$, respectively. The
spin dependent conductance $G^{\gamma}$ is given by
\begin{equation}
G^{\gamma}=g_{0}\sum_{j} T^{\gamma}_{j}\ ,
\end{equation}
where $g_{0}=2e^{2}/h$, factor $2$ is for valley degeneracy and summation goes over all of the modes. The
reflection and transmission probabilities for spin $\gamma$ carriers in the $j$-th mode is given by the
following relations
\begin{eqnarray}&&
R^{\gamma}_{j}=\frac{X'^{2}_{j}+Y'^{2}_{j}}{X_{j}^{2}+Y_{j}^{2}}\ ,\\&&
T^{\gamma}_{j}=\frac{4(k^{\gamma}R_{o})[z_{j,2}\xi_{j\uparrow}^{(1)}(R_{o})
\xi_{j\downarrow}^{(2)}(R_{o})-z_{j,1}\xi_{j\downarrow}^{(1)}(R_{o})
\xi_{j\uparrow}^{(2)}(R_{o})]}{k^{\gamma}R_{i}(X_{j}^{2}+Y_{j}^{2})}\ ,
\end{eqnarray}
Symbols in the above equations are defined as
\begin{eqnarray}\nonumber
X ^\prime _{j} = z_{j,1}z_{j,2}w^{-}_{j\downarrow \downarrow}-w^{-}_{j\uparrow \uparrow}\ , \quad Y^\prime
_{j}= z_{j,2}w^{-}_{j\uparrow \downarrow}+ z_{j,1}w^{-}_{j\downarrow \uparrow}\ ,\\\nonumber  X_{j}=w^{-}_{j
\uparrow \uparrow}+z_{j,1}z_{j,2}w^{-}_{j\downarrow \downarrow}\ , \quad Y_{j}=z_{j,2}w^{+}_{j \uparrow
\downarrow}- z_{j,1} w^{+}_{j\downarrow \uparrow}\ ,
\\ w^{\pm}_{js s^\prime}=\xi ^{(1)}_{js}(R_{i})
\xi ^{(2)}_{js^\prime}(R_{o}) \pm \xi ^{(1)}_{js}(R_{o}) \xi ^{(2)}_{js^\prime}(R_{i})\ .
\end{eqnarray}
Pumped current for spin $\gamma$ carriers (Eq. \ref{Brouwer-formula}) can be rewritten as the following
explicit form,
\begin{equation}
I^{\gamma}_{p}=I_{0}\sum_{j} Im\left( \frac{\partial r^{\gamma}_{L,L}}{\partial U} \frac{\partial
(r^{\gamma}_{L,L})^\ast}{\partial B} + \frac{\partial t^{\gamma}_{L,R}}{\partial U} \frac{\partial
(t^{\gamma}_{L,R})^\ast}{\partial B}\right)\ .
\end{equation}
In the next section we give the results for charge and spin conductances and pumped currents. We investigate
the effect of the magnetic filed and specially the Zeeman splitting on the transport properties of the
graphene Corbino pump.
%
%%%%%%%%%%%%%%%%%%%%%%%%%%%%%%%%%%%%%%%%%%%%%%%%%%%%%%%%%%%%%%%%%%%%
%%%%%%%%%%%%%%%%%%%%%      Fig 2ab   %%%%%%%%%%%%%%%%%%%%%%%%%%%%%%%
\begin{figure}
\centerline{\includegraphics[width=7cm]{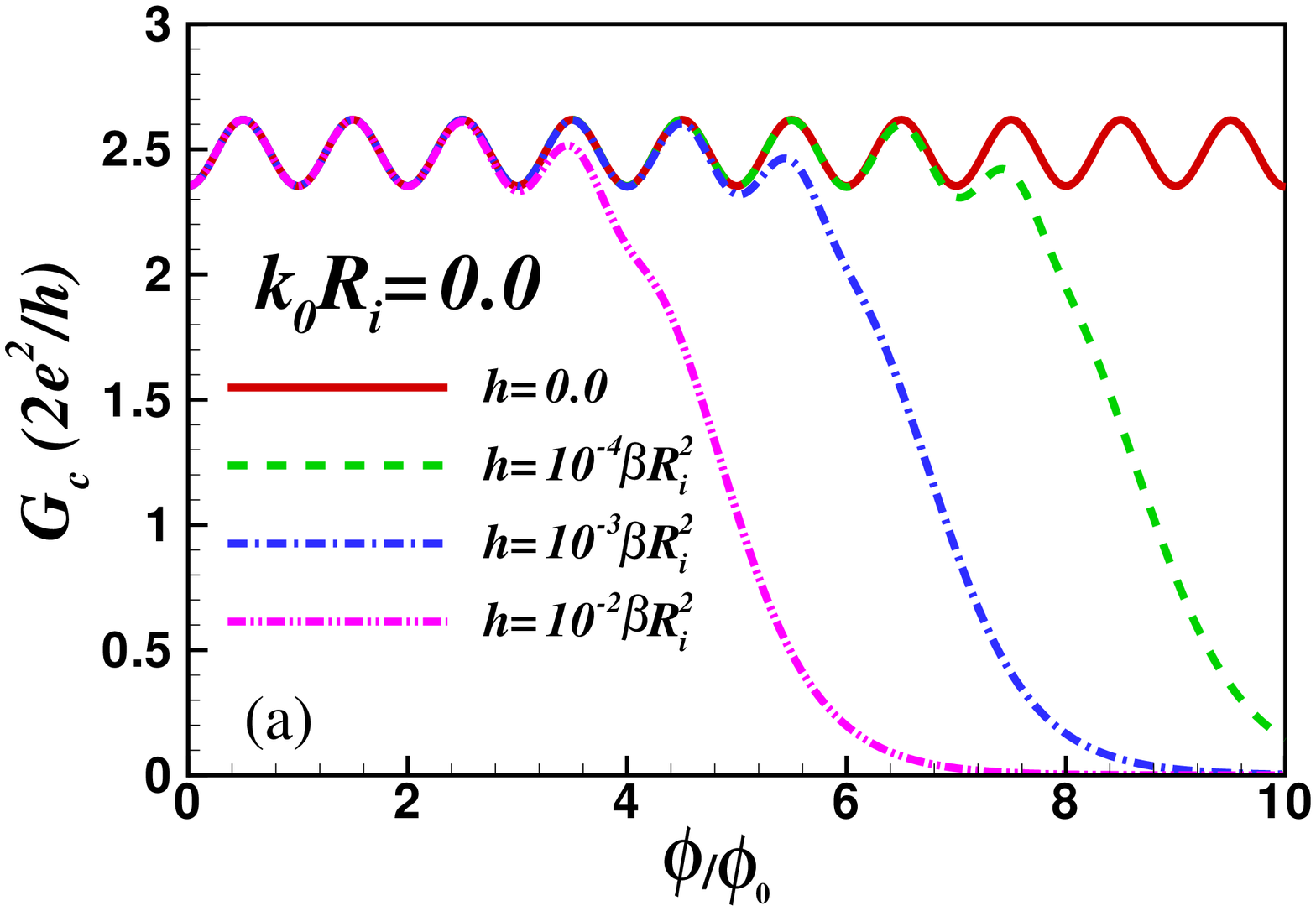}} \centerline{\includegraphics[width=7cm]{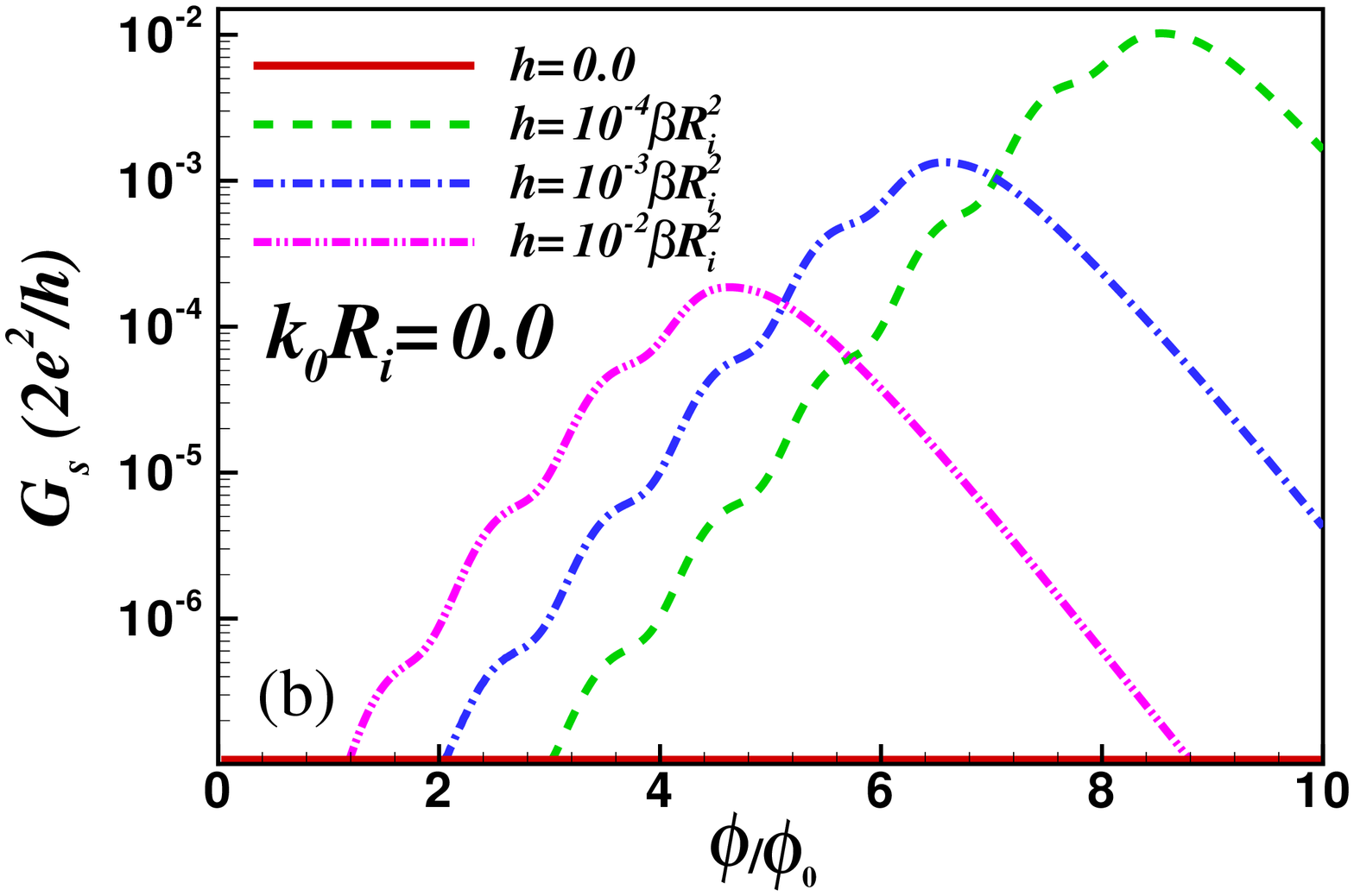}}
\caption{(Color online) (a) Charge conductance at the zero doping as a function of the magnetic flux piercing
the ring area and for different dependencies of the reduced Zeeman splitting $h=E_zR_i/\hbar v_F=\zeta\beta
R_i^2$ for $R_O/R_i=5$. Charge conductance is a periodic function of the magnetic flux at zero doping and in
the absence of the Zeeman splitting. The Zeeman splitting leads to suppression of the charge conductance. (b)
Spin conductance as a function of $\phi/\phi_0$ and for the same parameters. Spin conductance is considerable
around the values of the magnetic field which charge conductance starts to suppress.} \label{GTS-k0Ri=0}
\end{figure}
%%%%%%%%%%%%%%%%%%%%%%%%%%%%%%%%%%%%%%%%%%%%%%%%%%%%%%%%%%%%%%%%%%%%
%
%
%%%%%%%%%%%%%%%%%%%%%%%%%%%%%%%%%%%%%%%%%%%%%%%%%%%%%%%%%%%%%%%%%%%%
%%%%%%%%%%%%%%%%%%%%%      Fig 3ab  %%%%%%%%%%%%%%%%%%%%%%%%%%%%%%%
\begin{figure}
\centerline{\includegraphics[width=7cm]{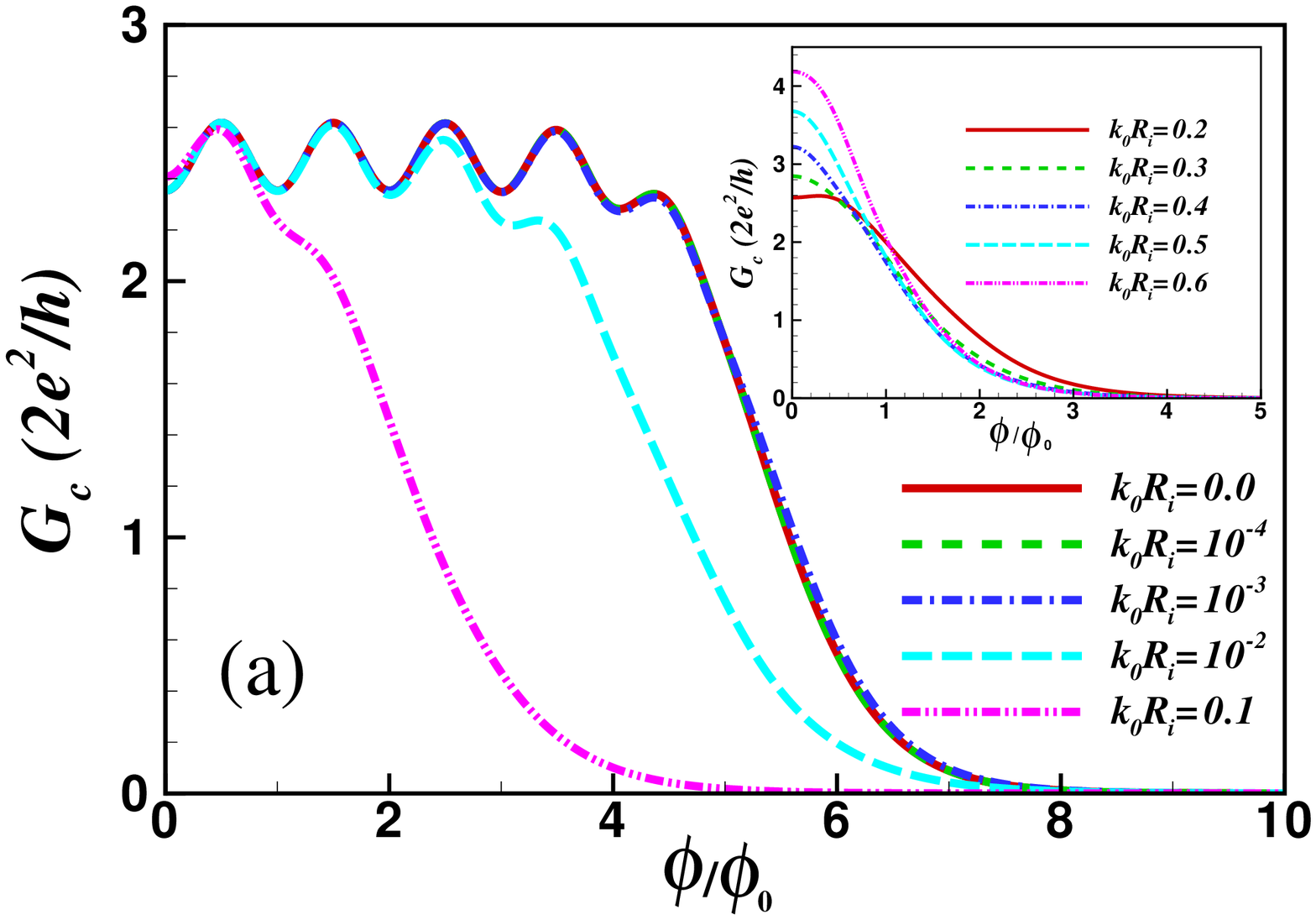}} \centerline{\includegraphics[width=7cm]{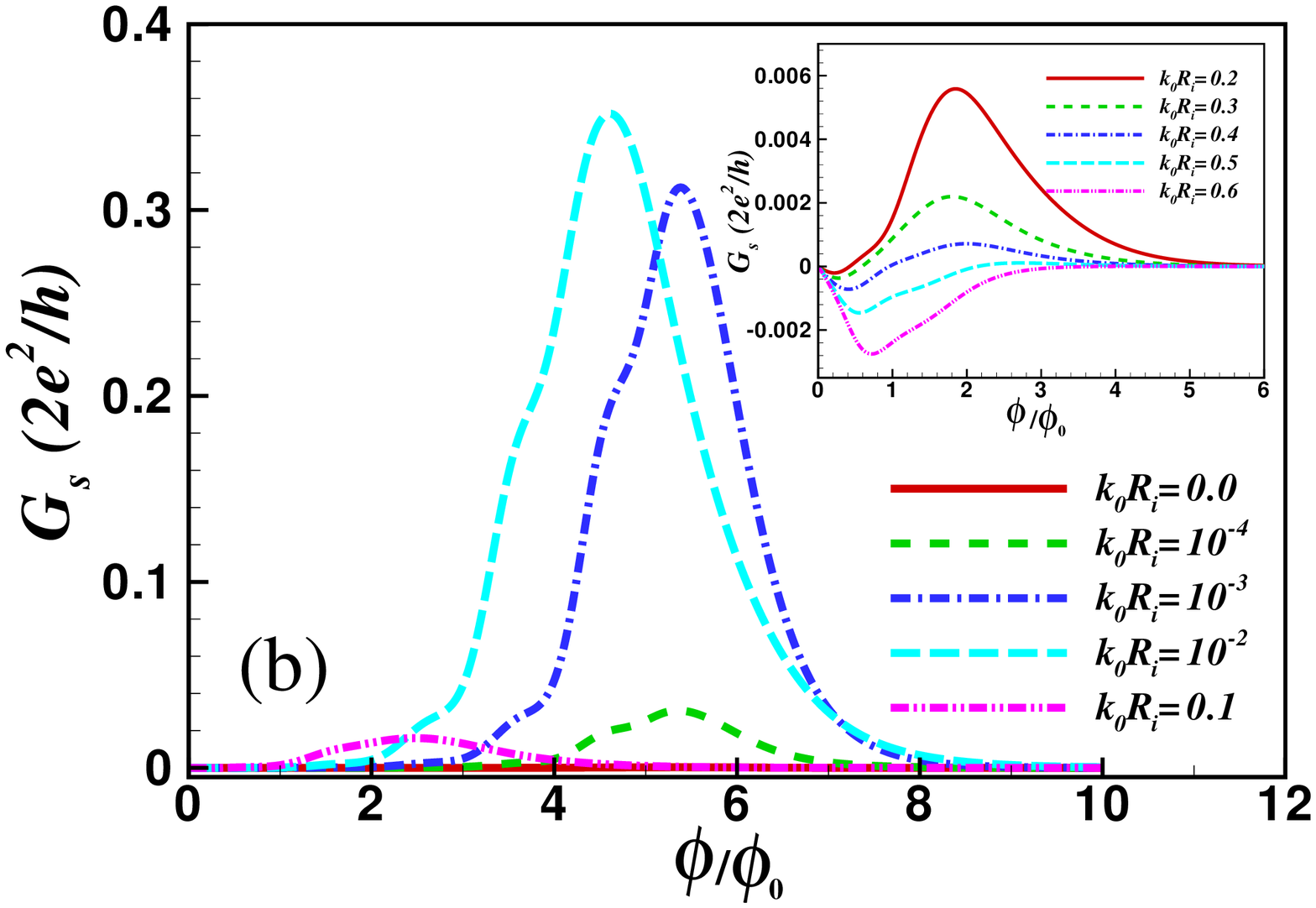}}
\caption{(Color online) (a) Charge conductance as a function of the magnetic flux piercing the ring area and
for different doping values for $R_i=20nm$ and $R_O/R_i=5$. Charge conductance starts to suppress at lower
values of the flux by increasing of the doping. (b) Spin conductance as a function of $\phi/\phi_0$ and for
the same parameters. Spin conductance is accompanied by the charge conductance just around where the later
starts to suppress. Insets of the figures show the behavior of the charge and spin conductances at large
dopings.} \label{GTS-Phi}
\end{figure}
%%%%%%%%%%%%%%%%%%%%%%%%%%%%%%%%%%%%%%%%%%%%%%%%%%%%%%%%%%%%%%%%%%%%
%
%
%%%%%%%%%%%%%%%%%%%%%%%%%%%%%%%%%%%%%%%%%%%%%%%%%%%%%%%%%%%%%%%%%%%%
%%%%%%%%%%%%%%%%%%%%%      Fig 4ab  %%%%%%%%%%%%%%%%%%%%%%%%%%%%%%%
\begin{figure}
\centerline{\includegraphics[width=7cm]{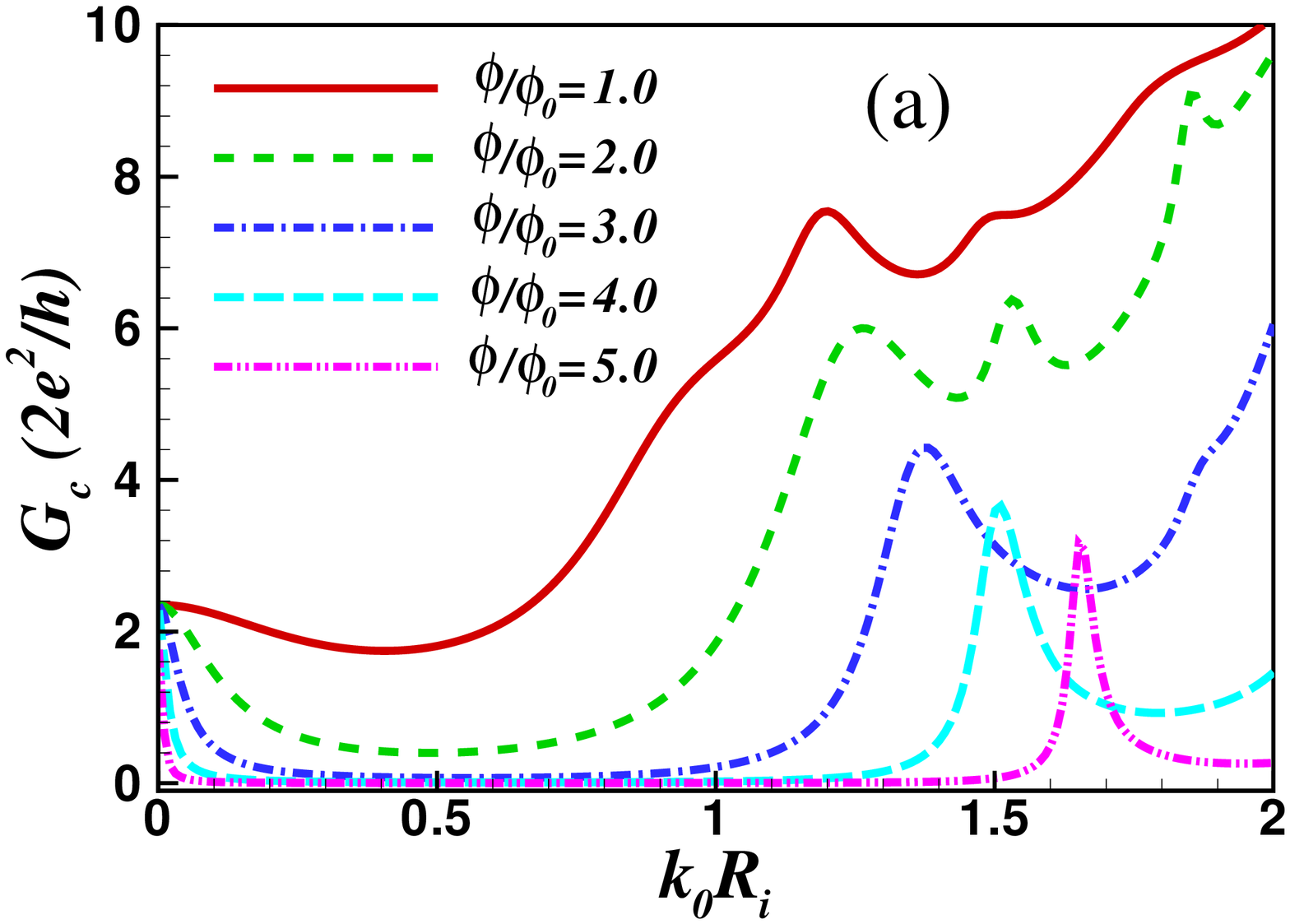}} \centerline{\includegraphics[width=7cm]{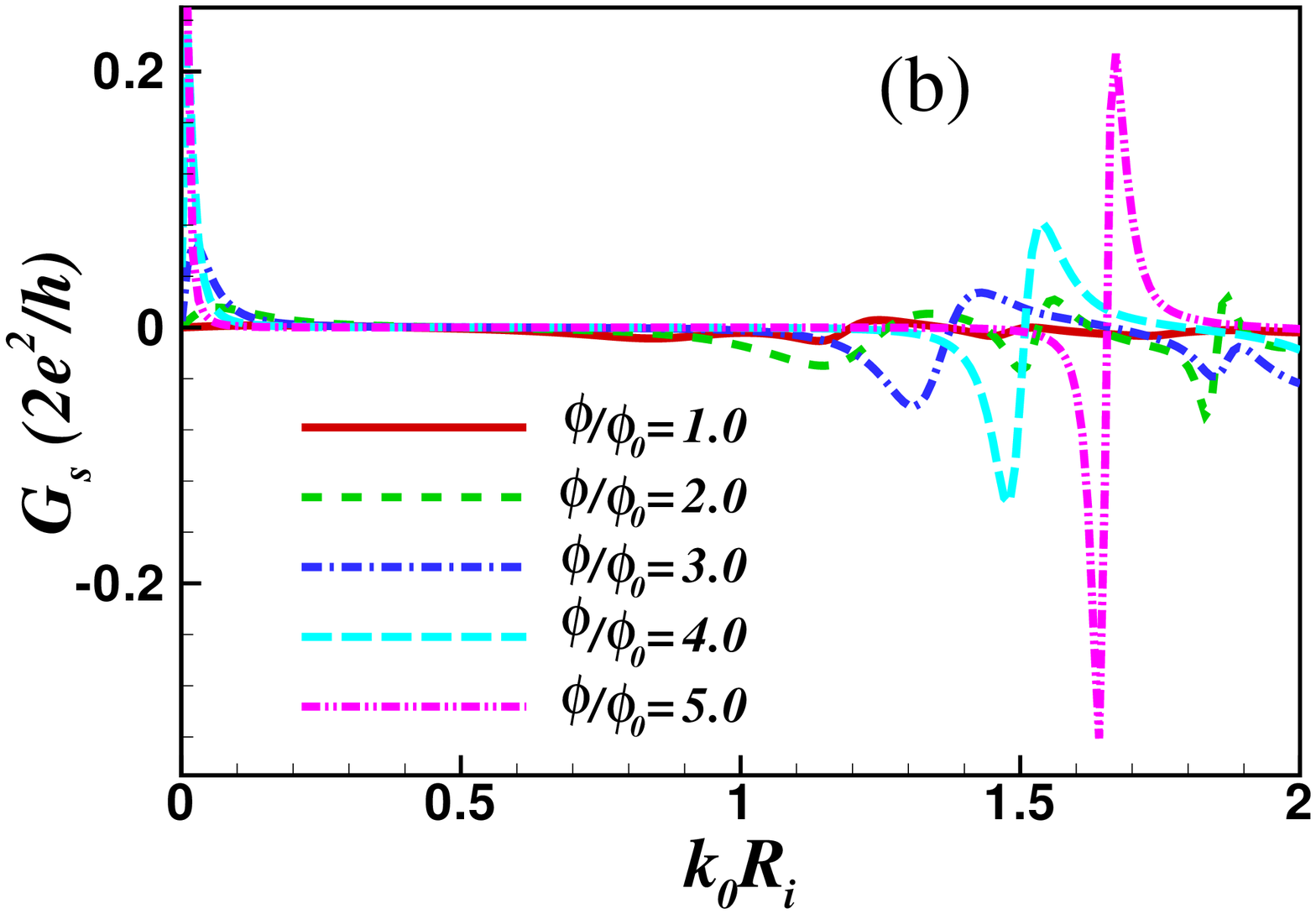}}
\caption{(Color online) (a) Charge conductance as a function of the doping and for different magnetic fluxes
piercing the ring area for $R_i=20nm$ and $R_O/R_i=5$. It shows crossover from the ballistic transport regime
to the field suppressed one by increasing $\phi$. (b) Spin conductance as a function of $k_0R_i$ and for the
same parameters. Spin conductance is nonzero just around where the charge conductance has peaks due to the
resonance with Landaue levels.} \label{GTS-k0Ri}
\end{figure}
%%%%%%%%%%%%%%%%%%%%%%%%%%%%%%%%%%%%%%%%%%%%%%%%%%%%%%%%%%%%%%%%%%%%
%
%
%%%%%%%%%%%%%%%%%%%%%%%%%%%%%%%%%%%%%%%%%%%%%%%%%%%%%%%%%%%%%%%%%%%%
%%%%%%%%%%%%%%%%%%%%%      Fig 5ab  %%%%%%%%%%%%%%%%%%%%%%%%%%%%%%%
\begin{figure}
\centerline{\includegraphics[width=7cm]{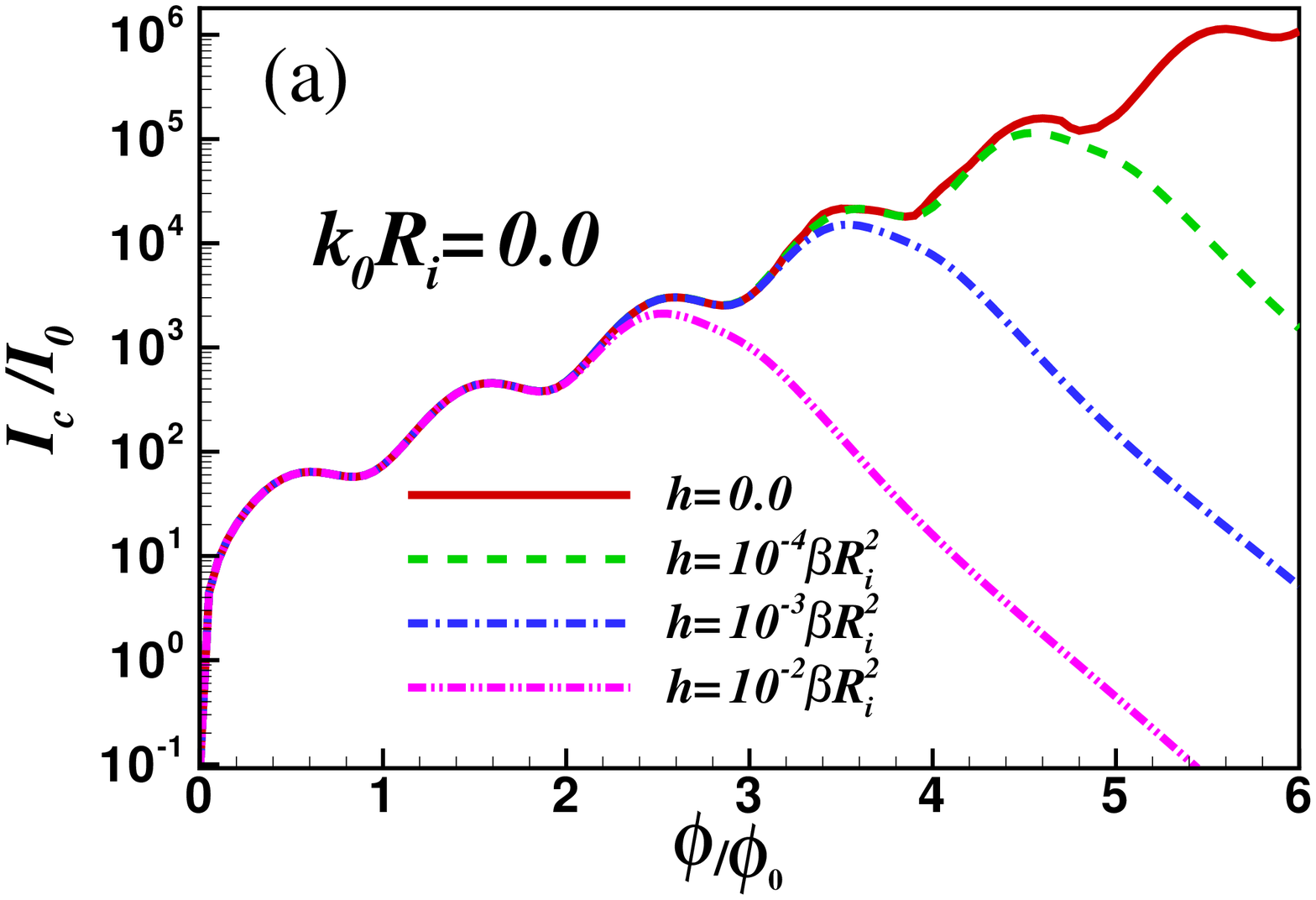}} \centerline{\includegraphics[width=7cm]{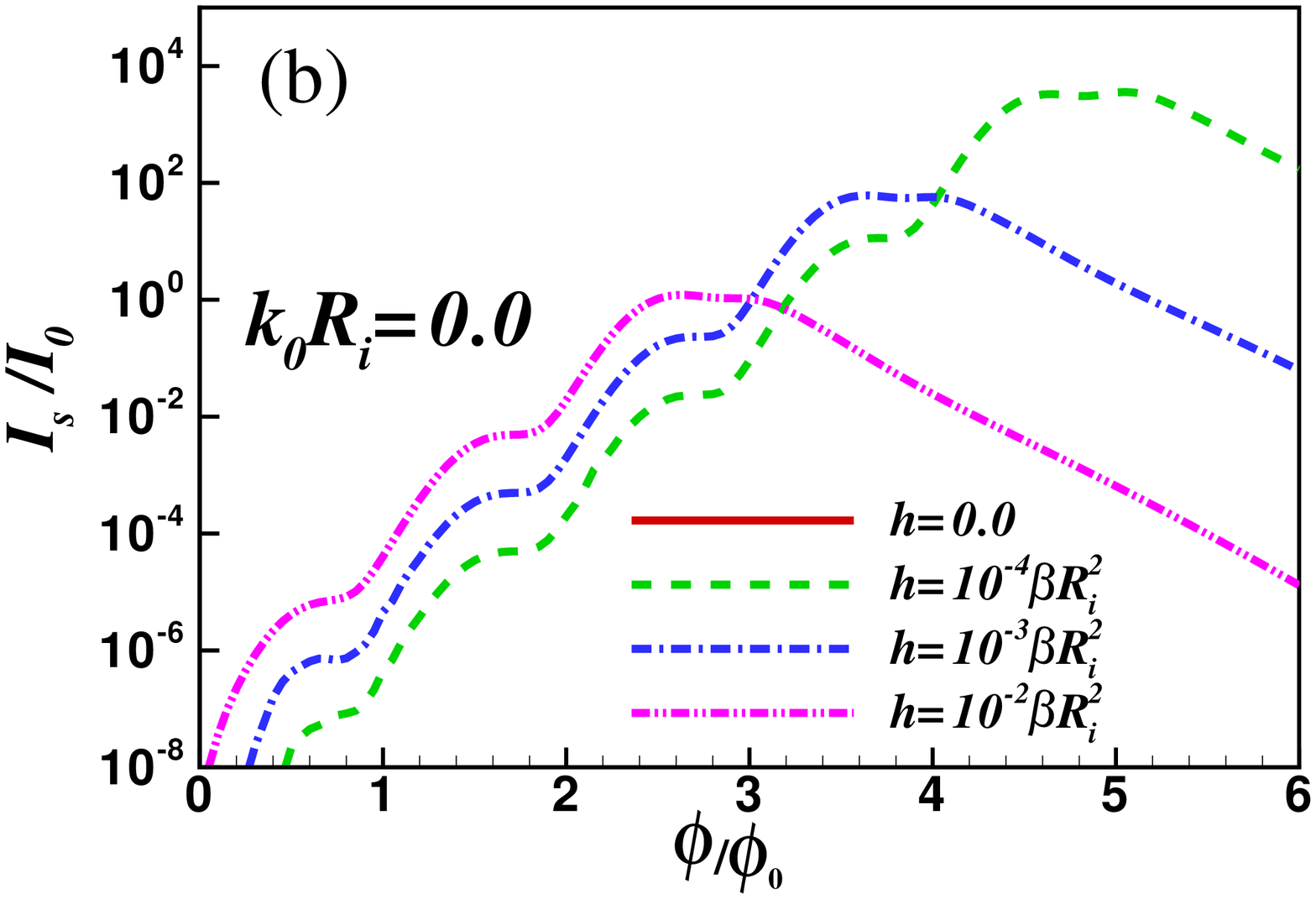}}
\caption{(Color online) (a) Pumped charge current at the zero doping as a function of the magnetic flux
piercing the ring area and for different dependencies of the reduced Zeeman splitting $h=E_zR_i/\hbar
v_F=\zeta\beta R_i^2$, and for $R_O/R_i=10$ in the logarithmic scale. It shows an increasing behavior as a
function of $\phi$ with a small oscillations. Inclusion of the Zeeman splitting leads to suppression of the
pumped charge current. (b) Pumped spin current as a function of $\phi/\phi_0$ and for the same parameters.
Pumped spin current is considerable around the values of the magnetic field where charge conductance starts to
suppress.} \label{IPTS-k0Ri=0}
\end{figure}
%%%%%%%%%%%%%%%%%%%%%%%%%%%%%%%%%%%%%%%%%%%%%%%%%%%%%%%%%%%%%%%%%%%%
%
%
%%%%%%%%%%%%%%%%%%%%%%%%%%%%%%%%%%%%%%%%%%%%%%%%%%%%%%%%%%%%%%%%%%%%%%%%%%%%%%%%%%%%%%%%%      Fig 6ab  %%%%%%%%%%%%%%%%%%%%%%%%%%%%%%%
\begin{figure}[h]
\centerline{\includegraphics[width=7cm]{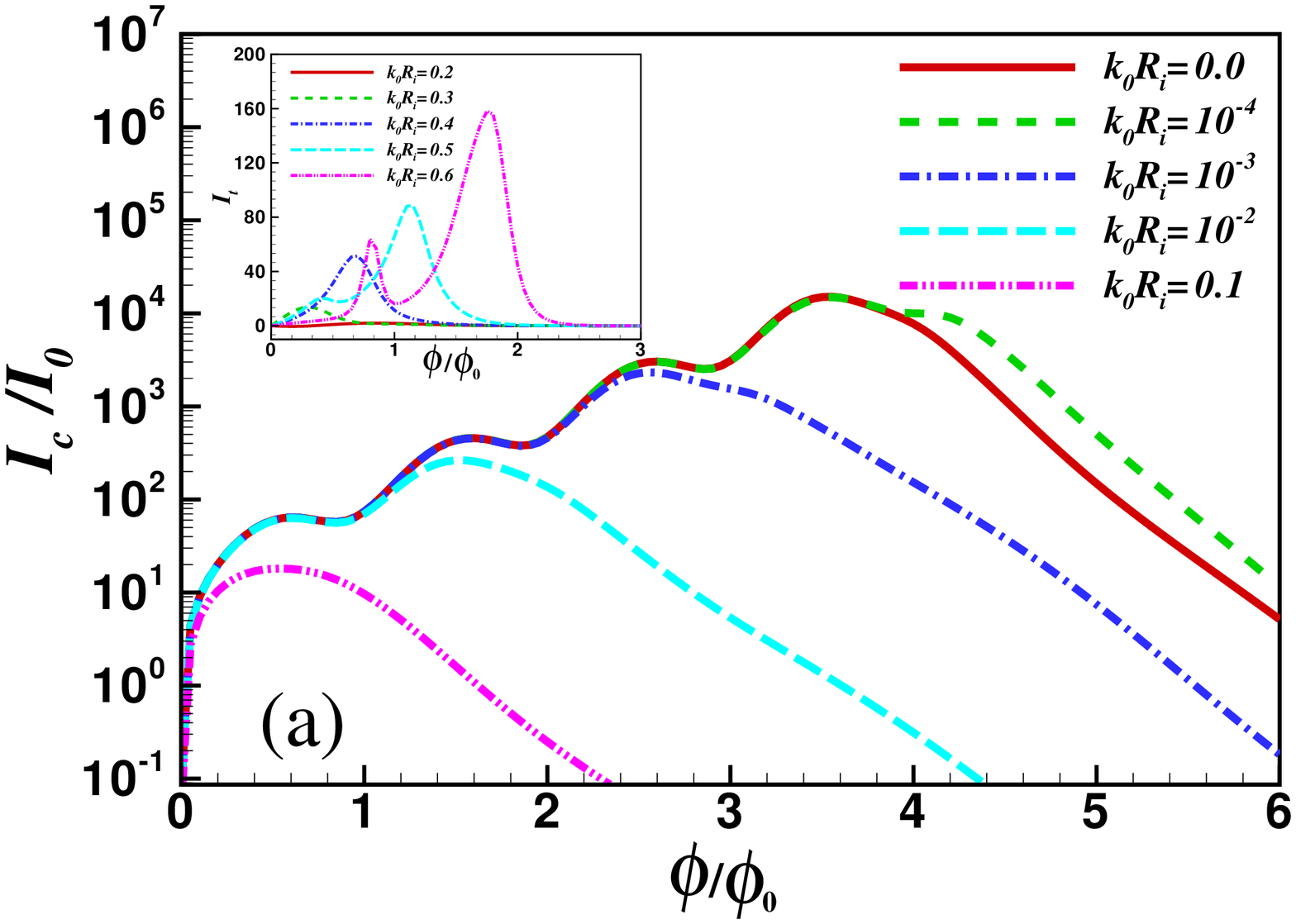}} \centerline{\includegraphics[width=7cm]{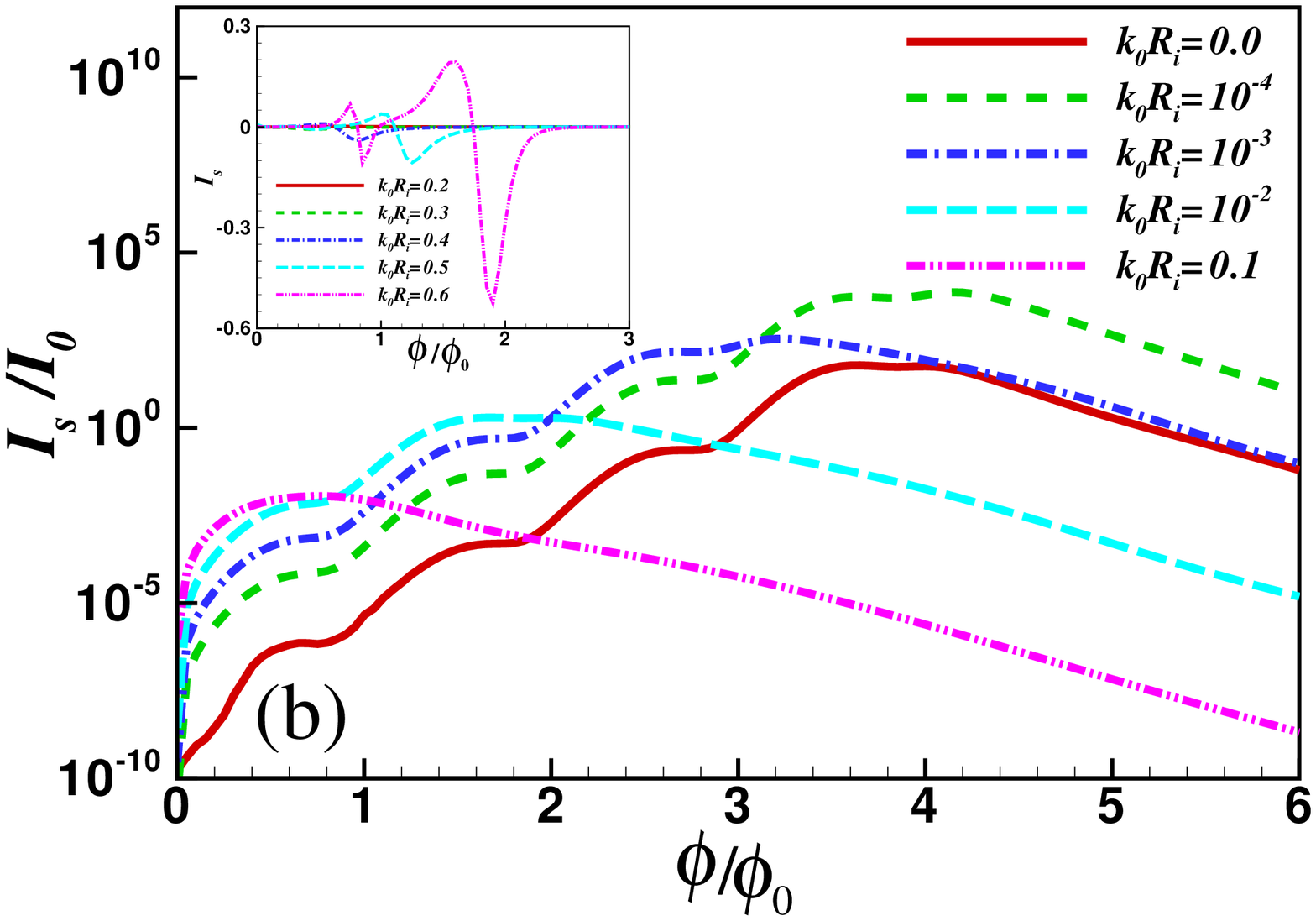}}
\caption{(Color online) (a) Pumped charge current as a function of the magnetic flux piercing the ring area
and for different dopings for $R_i=50nm$ and $R_O/R_i=10$ in the logarithmic scale. Pumped charge current
starts to suppress at lower values of the flux by increasing of the doping. (b) Pumped spin current as a
function of $\phi/\phi_0$ and for the same parameters. Pumped spin current has a very similar behavior with
pumped charge current. Insets of the figures show the behavior of the pumped charge and spin currents at large
doping values.} \label{IPTS-phi}
\end{figure}
%%%%%%%%%%%%%%%%%%%%%%%%%%%%%%%%%%%%%%%%%%%%%%%%%%%%%%%%%%%%%%%%%%%%
%
%%%%%%%%%%%%%%%%%%%%%%%%%%%%%%%%%%%%%%%%%%%%%%%%%%%%%%%%%%%%%%%%%%%%
%%%%%%%%%%%%%%%%%%%%%      Fig 7  %%%%%%%%%%%%%%%%%%%%%%%%%%%%%%%
\begin{figure}[t]
\centerline{\includegraphics[width=7cm]{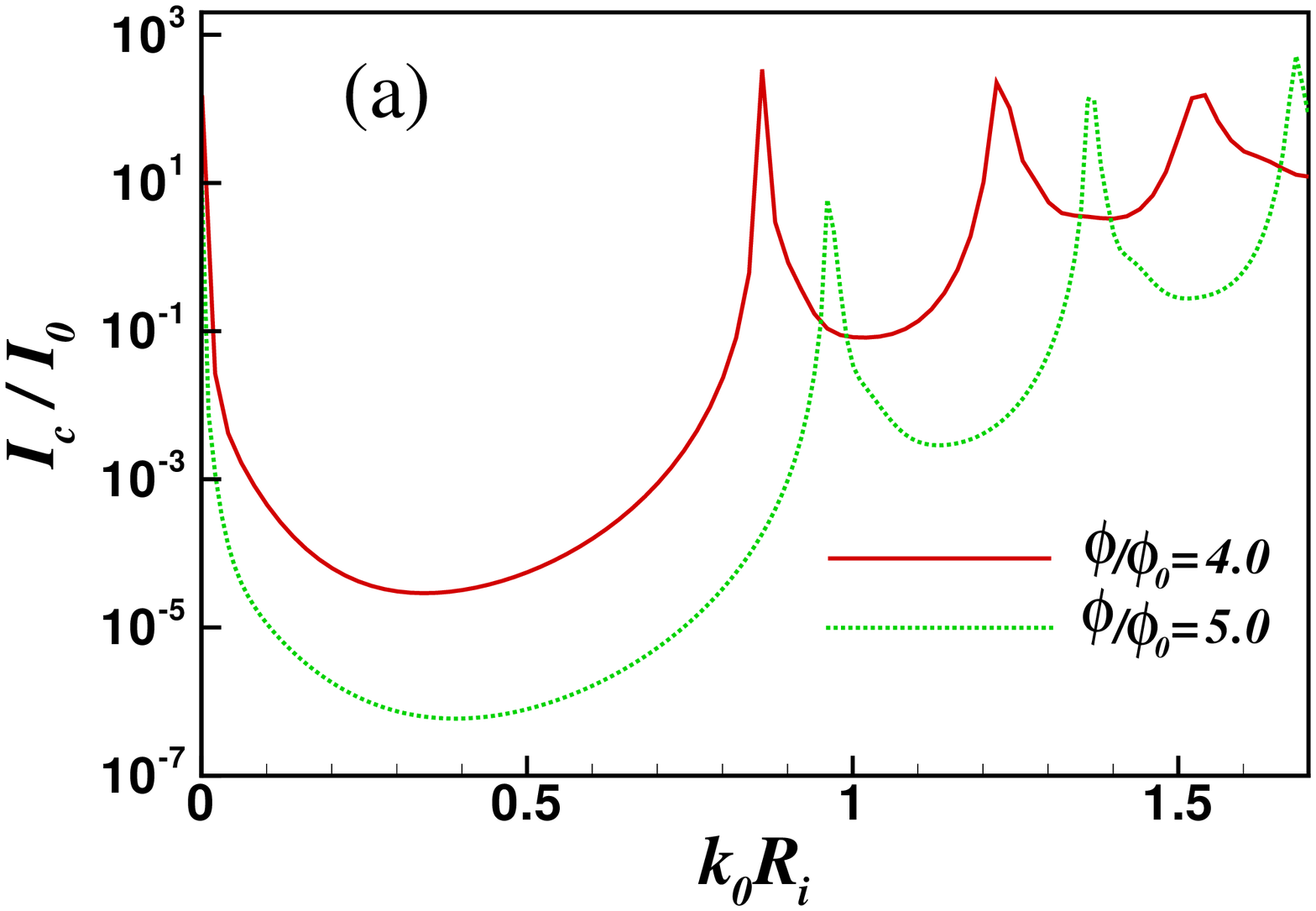}} \centerline{\includegraphics[width=7cm]{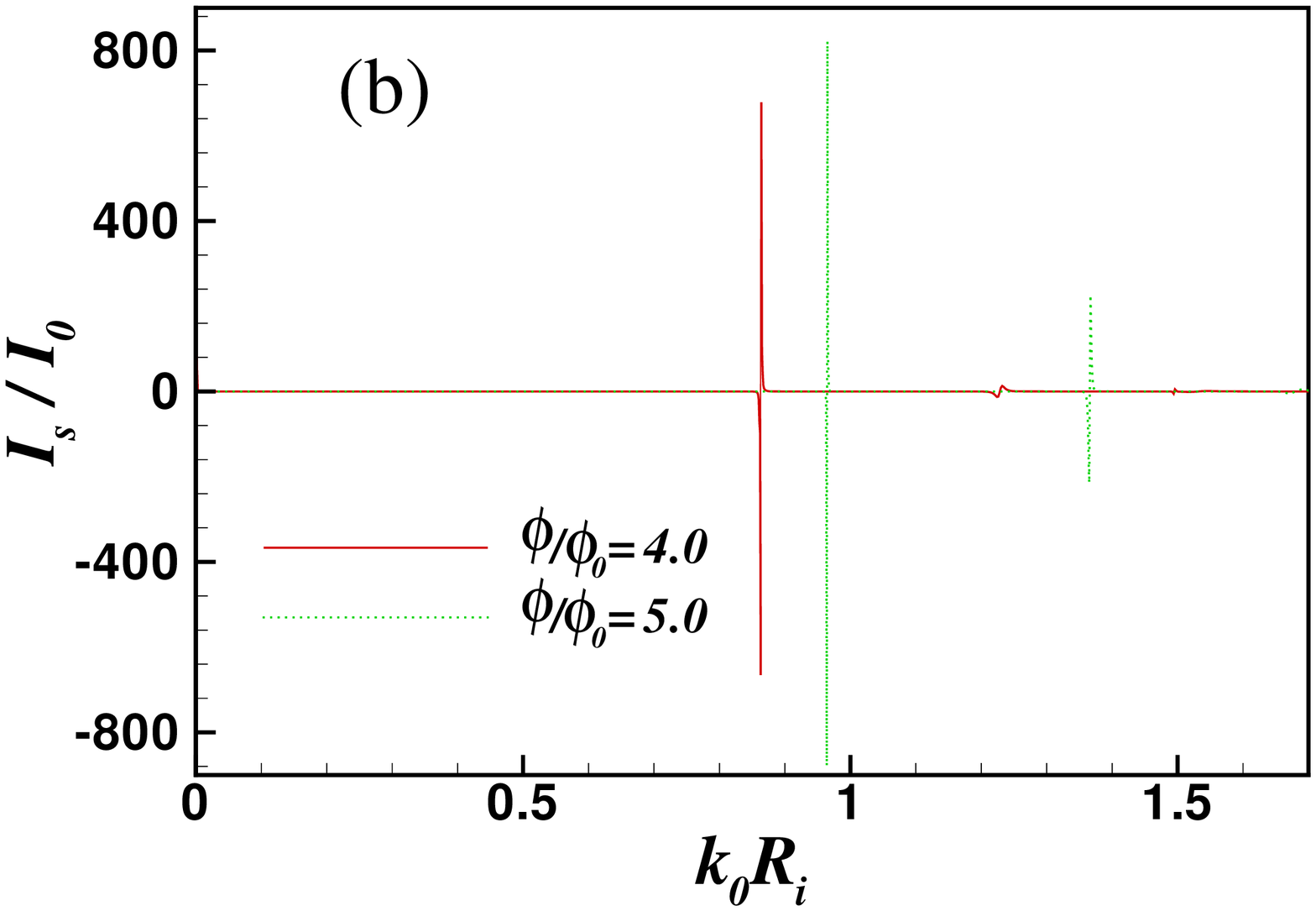}}
\caption{(Color online) Pumped charge current as a function of the doping and for different magnetic fluxes
piercing the ring area for $R_i=50nm$ and $R_O/R_i=10$ in the logarithmic scale. Similar to the charge
conductance it has peaks at resonances with Landaue levels in the field suppressed regime. (b) Spin pumped
current as a function of $k_0R_i$ and for the same parameters. Spin pumped current is nonzero just around
where the charge conductance has peaks due to the resonance with Landaue levels.} \label{IPTS-k0Ri}
\end{figure}
%%%%%%%%%%%%%%%%%%%%%%%%%%%%%%%%%%%%%%%%%%%%%%%%%%%%%%%%%%%%%%%%%%%%
%

%%%%%%%%%%%%%%%%%%%%%%%%%%%%%%%%%%%%%%%%%%%%%%%%%%%%%%%%%%%%%%%%%%%%
\section{Results and Discussions}\label{results}
%%%%%%%%%%%%%%%%%%%%%%%%%%%%%%%%%%%%%%%%%%%%%%%%%%%%%%%%%%%%%%%%%%%%

Let us first study the effect of the Zeeman splitting on the spin and charge conductances. We start with
analyzing the conductance at the Dirac point. As it has been shown in the Ref. \cite{Rycerz10}, at the Dirac
point conductance shows a nearly periodic oscillations in terms of the magnetic flux piercing the ring and
indeed it is independent of the magnetic flux piercing the inner lead. In Fig. \ref{GTS-k0Ri=0}(a), the Zeeman
splitting ($h=E_zR_i/\hbar v_F=\zeta\beta R^2_{i}$ with $\zeta=\hbar/m_ev_FR_i$ and $m_e$ the electron mass)
leads to dependence of the conductance on the magnetic flux piercing the inner lead and it causes suppression
of the charge conductance when $\phi$ exceeds the flux threshold value $\phi/\phi_0\sim2\ln(\zeta)$. In the
lower magnetic fields, curves for nonzero Zeeman splitting follow the zero one. Fig. \ref{GTS-k0Ri=0}(b) shows
the spin conductance at the Dirac point, in the logarithmic scale, as a function of $\phi/\phi_0$  for
different values of $\zeta$. Spin conductance is negligible for all values of $\phi$ except for the values
around the flux threshold value. We continue by investigating the charge and spin conductances out of the
Dirac point. Figs. \ref{GTS-Phi} (a) and (b) show the charge and spin conductances as a function of $\phi$ for
different doping values $k_0R_i=|E-U_0|R_i/\hbar v_F$ differing one order of the magnitudes, respectively.
Insets of the figures show the results for large values of $k_0R_i$. Suppression of the charge conductance for
nonzero doping starts at the threshold value given by,
\begin{equation}
\phi/\phi_0\sim\min\{2\log(\zeta),2\log(k_0R_i)\}\ . \label{threshold-flux}
\end{equation}
Thus, Zeeman splitting and doping compete with each other for suppressing the charge conductance. Despite of
the smallness of the Zeeman splitting it produces a considerable spin conductances, about one order of
magnitude smaller than the charge conductance, around the threshold value given by Eq. (\ref{threshold-flux})
for small nonzero doping values. For large doping it is almost three order of magnitude smaller than the
charge conductance and thus it is negligible.

Results for charge and spin conductances as a function of the doping and for different values of the magnetic
field have been shown in Figs. \ref{GTS-k0Ri} (a) and (b), respectively. These plots show crossover from the
low to high magnetic fields regimes. It is accompanied by the crossover from {\it ballistic transport regime}
to {\it field-suppressed regime} which is characterized by the condition $2R_c=R_o-R_i$, with
$R_c=(k^{+}+k^{-})l_B^2/2$. At high magnetic fields satisfying the condition
\begin{equation}
\frac{\phi}{\phi_0}>\frac{(R_o/R_i-1)}{4\ln(R_o/R_i)}(k^{+}+k^{-})R_i\ , \label{critical-flux}
\end{equation}
charge conductance is suppressed except from the isolated peaks which correspond to the resonances with Landau
levels. The spin conductance gets considerable only in the filed suppressed regime characterized by the above
equation and at the resonance condition. Further, upon the total conductance shows a peak, the spin
conductance changes sign through a singularity.

To study magneto pumping current we plot the pumped charge and spin currents as a function of the flux
piercing the ring and doping. In Figs. \ref{IPTS-k0Ri=0} (a) and (b) we give the pumped charge and spin
currents against the $\phi/\phi_0$ for different values of $\zeta$ at zero doping ($k_0R_i=0$) and
$R_o/R_i=10$. In the absence of the Zeeman splitting the pumped charge current shows an enhancement in the
logarithmic scale as a function of $\phi/\phi_0$ and with small oscillations having a period identical to the
conductance oscillations period. It is similar to the behavior of the conductance at zero doping except that
the pumped charge current increases with increasing the flux, whereas the charge conductance oscillates around
a fixed value. Zeeman splitting has a similar effect on the pumped charge and spin currents with those of the
charge and spin conductances. It suppresses the pumped charge and spin currents around the flux threshold
value given by $\phi/\phi_0\sim2\ln(\zeta)$. The pumped spin current is almost two order of magnitude smaller
than the pumped charge current and it has only considerable values for $\phi/\phi_0$ smaller and close to the
flux threshold value. In Figs. \ref{IPTS-phi} (a) and (b) we compare the pumped charge and spin currents for
the nonzero doping values with those of the zero one. The insets of the figures show the results in the high
doping limit. Results for the pumped current show similar behavior with the conductance. They continue to
increase as a function of the flux, with small oscillations, until they are suppressed by the effects of the
Zeeman splitting or doping around the flux threshold value given by Eq. \ref{threshold-flux}. In the limit of
the large doping the pumped charge current shows some peaks as the flux increases until it inters to the {\it
field-suppressed regime} characterized by the flux values given by Eq. \ref{critical-flux}. At the same time
the pumped spin current changes sign when the pumped charge current shows a peak. Values of the pumped charge
and spin currents in the high doping limit are very less than their values in the low doping. In Figs.
\ref{IPTS-k0Ri} (a) and (b) we show results for pumped charge and spin currents as a function of the doping
and for high values of the magnetic field. As it is apparent from the figures both of them has considerable
values only when a resonance occurs due to the Landau levels. When charge current shows a peak the spin
current changes sign through a discontinuity. The pumped spin current have quite large values at the
resonances and its accompaniment with the sign change makes it highly probable to be detected in the
experiment. This result may be very interesting for practical usage in graphene spintronics.

About the feasibility of the experimental observation of the effects studied here we can address the charge
transport measurement which recently has been done in dual gated bilayer graphene with Corbino
geometry\cite{Yan10}. Moreover, magnetoresistance measurements in tilted magnetic fields have been done in
graphene with planar and Corbino geometries as well\cite{Zhao12}. Also, recently quantum pumping in graphene
has been realized in the graphene double quantum dot pump. In this experiment the pumped current is driven by
two oscillating gate potentials operating on the total number of the electrons on each quantum dot. In the
graphene Corbino pump introduced here, values of the pumped charge and spin currents are in the experimental
range and can be detected easily. To estimate these values we consider typical values for the experimental
parameters $\delta U/U=0.01$ and $\delta B/B=0.01$ and $\omega/2\pi=10$ GHz. The resulting pumped charge and
spin currents will be in the ranges of the $\sim 1$ nA and $\sim 0.1$ nA, respectively. These values of the
pumped charge and spin currents are in the experimentally observable range.

%%%%%%%%%%%%%%%%%%%%%%%%%%%%%%%%%%%%%%%%%%%%%%%%%%%%%%%%%%%%%%%%%%%%
\section{Conclusion}\label{conclusion}
%%%%%%%%%%%%%%%%%%%%%%%%%%%%%%%%%%%%%%%%%%%%%%%%%%%%%%%%%%%%%%%%%%%%
In conclusion we have considered a graphene based quantum pump with Corbino geometry which possesses the
advantage of the absence of edges. The pump is driven by an electrostatic potential applied to the ring region
and a magnetic field applied perpendicular to the graphene. We studied the effect of the Zeeman splitting on
the charge and spin conductances and pumped currents. The Zeeman splitting, despite of its smallness, has a
considerable effect on the charge conductance and pumped current and it generates a considerable spin
conductance and pumped spin current. It suppresses the periodic oscillations of the conductances and pumped
currents at the zero doping. Also, it produces a considerable spin conductance and pumped spin current at the
values of the magnetic field which charge conductance and pumped charge current start to suppress,
respectively. Moreover, the pumped charge and spin currents grow up by increasing the magnetic filed until
they are suppressed due to the effects of the doping and the Zeeman splitting. It let us control the magnitude
of the pumped charge and spin currents by the doping level and magnetic field in such a way that we can have
quite considerable values of these pumped currents at the Dirac point and higher doping levels.

%%%%%%%%%%%%%%%%%%%%%%%%%%%%%%%%%%%%%%%%%%%%%%%%%%%%%%%%%%%%%%%%%%%%%%
%%%%%%%%%%%%%%%%%%%%%%%%%%%%%%%%%%%%%%%%%%%%%%%%%%%%%%%%%%%%%%%%%%%%%%

%\begin{acknowledgments}
%
%\end{acknowledgments}

\section*{References}
\end{document}